\documentclass[twocolumn,showpacs,preprintnumbers,prl,amsmath,amssymb]{revtex4}

\usepackage{bm}

\begin{document}

\title{Analytic Coulomb matrix elements in a three-dimensional geometry}

\author{Jaime \surname{Zaratiegui Garc{\'\i}a}}
 \affiliation{Department of Physical Sciences, P.O. Box 3000,
 FIN-90014 University of Oulu, Finland} %
 \email{jaime.zaratiegui@oulu.fi}

\date{\today}

\begin{abstract}
Using a complete basis set we have obtained an analytic expression for
the matrix elements of the Coulomb interaction. These matrix elements
are written in a closed form. We have used the basis set of the
three-dimensional isotropic quantum harmonic oscillator in order to
develop our calculations, which can be useful when treating
interactions in localized systems.
\end{abstract}


\maketitle

\section{\label{sec:intro}Introduction}

Having an analytic expression for the Coulomb matrix elements is an
important step for several numerical methods, like, for example, exact
diagonalization method.
In order to describe the Coulomb interaction in three dimensions, we
have chosen the basis set of the isotropic harmonic oscillator for the
single-particle wave functions, which, in one dimension, is written as
\begin{equation}
\psi_{n_x}(x) = \left(a\sqrt{\pi}2^{n_x}n_x!\right)^{-1/2}
e^{-\frac{1}{2}x^2/a^2} H_{n_x}(x/a), \label{eq:wf}
\end{equation}
where $a=\sqrt{\hbar/m\omega}$ is taken as the characteristic unit
length. One of the reasons for the election of this particular
basis set is the Gaussian Product Theorem, which guarantees that the
product of two Gaussian type orbitals (a linear combination of them in
our case) centered on two different atoms is a finite sum 
of Gaussians centered on a point along the axis connecting
them.

In previous works, several ways to evaluate the two-dimensional matrix
elements using different approaches have been
studied~\cite{halonen,stone,b:tapash}, such as restricting to the
lowest Landau level due to simplicity reasons~\cite{tsiper,girvin}.

The purpose of this paper is to report an analytic formula for
the Coulomb interaction written in closed form. It can be easily
implemented by computer means and could help to improve the
performance of solid state simulations in which interactions are taken
into account.

\section{\label{sec:elem}Matrix elements}

In order to derive an analytical expression for the Coulomb
interaction matrix elements we will proceed starting with the same
approach as the one used in Ref. \cite{chakra}, i.e., writing the
single-electron wave function and the Coulomb potential as their
Fourier transform integrals:
\begin{eqnarray}
\psi_\lambda(\bm{r}) & = & \frac{1}{(2\pi)^{3/2}}\int
\phi_\lambda(\bm{q}) e^{-i\bm{q}\cdot\bm{r}}\,\mathrm{d}\bm{q}, \\
V(\bm{r}) & = & \frac{1}{(2\pi)^{3/2}}\int \tilde{V}(\bm{q})
e^{-i\bm{q}\cdot\bm{r}}\,\mathrm{d}\bm{q},
\end{eqnarray}
where $\lambda$ stands for a set of quantum numbers
$\left\{n_i\right\}$ and $V(\bm{r_1}-\bm{r_2})=r_{12}^{-1}$ is the
Coulomb potential. Now, the two-particle matrix element, which, in
real space is written as
\begin{eqnarray}
\mathcal{V}^{\lambda_1\lambda_2}_{\lambda_3\lambda_4}  = \int
\psi^\ast_{\lambda_1}(\bm{r}_1) \psi^\ast_{\lambda_2}(\bm{r}_2)
V(\bm{r}_1-\bm{r}_2) \nonumber \\
 \times \psi_{\lambda_3}(\bm{r}_2) \psi_{\lambda_4}(\bm{r}_1)\,
\mathrm{d}\bm{r}_1\bm{r}_2,
\end{eqnarray}
is now expressed, in momentum space, as
\begin{eqnarray}
\mathcal{V}^{\lambda_1\lambda_2}_{\lambda_3\lambda_4} =
\frac{1}{(2\pi)^{3/2}} \int \phi^\ast_{\lambda_1}(\bm{q}_1)
\phi_{\lambda_4}(\bm{q}_1-\bm{q})\nonumber \\
\times \phi^\ast_{\lambda_2}(\bm{q}_2) \phi_{\lambda_3}(\bm{q}_2+\bm{q})
\tilde{V}(\bm{q})\,\mathrm{d}\bm{q}_1\mathrm{d}\bm{q}_2\mathrm{d}\bm{q}.
\label{eq:me2}
\end{eqnarray}
Eq.~(\ref{eq:me2}) can be rewritten in a more convenient and compact
form. Let us define $C^{\lambda}_{\lambda^\prime}(\bm{q})$ and
$D^{\lambda}_{\lambda^\prime}(\bm{q})$ as the following convolution
integrals:
\begin{eqnarray}
C^{\lambda}_{\lambda^\prime}(\bm{q}) & = & \int\phi_\lambda^\ast(\bm{k})
 \phi_{\lambda^\prime}(\bm{k}-\bm{q})\,\mathrm{d}\bm{k} \label{eq:C1}
 \\
 & = & \int\psi_\lambda^\ast(\bm{r})\psi_{\lambda^\prime}
 (\bm{r})e^{-i\bm{q}\cdot\bm{r}}\,\mathrm{d}\bm{r}, 
 \label{eq:C2} \\
D^{\lambda}_{\lambda^\prime}(\bm{q}) & = &
 \int\phi_\lambda^\ast(\bm{k})\phi_{\lambda^\prime} (\bm{k}+\bm{q})\,
 \mathrm{d}\bm{k} \label{eq:D1} \\ 
 & = & \int\psi_\lambda^\ast(\bm{r})\psi_{\lambda^\prime}
 (\bm{r})e^{i\bm{q}\cdot\bm{r}}\, 
 \mathrm{d}\bm{r} \label{eq:D2} \\ 
 & = & C^{\lambda}_{\lambda^\prime}(-\bm{q}).\label{eq:D3}
\end{eqnarray}
Substituting Eqs.~(\ref{eq:C1}) and (\ref{eq:D1}) into
Eq.~(\ref{eq:me2}) we obtain
\begin{equation}
\mathcal{V}^{\lambda_1\lambda_2}_{\lambda_3\lambda_4} =
\frac{1}{(2\pi)^{3/2}} \int C^{\lambda_1}_{\lambda_4}(\bm{q})
D^{\lambda_2}_{\lambda_3}(\bm{q}) \tilde{V}(\bm{q})\,\mathrm{d}\bm{q}.
\end{equation}

Now, it is straightforward to perform the integral appearing in
Eq.~(\ref{eq:C2})~\cite{gradshteyn}. Using Cartesian coordinates, it
is possible to separate all three variables and integrate
independently. For simplicity reasons, let us integrate only along the
$x$ variable, the result then reads:
\begin{eqnarray}
C^{n_x^1}_{n_x^4}(q_x) = \left(\frac{2^{n_{x+}^{14}}}{n_{x+}^{14}!}
\frac{n_{x-}^{14}!}{2^{n_{x-}^{14}}} \right)^{1/2} i^{n_x^1+n_x^4}
(-1)^{n_{x+}^{14}} \nonumber \\
\times e^{-q_x^2a^2/4} \left( \frac{aq_x}{2} \right)^{|n_x^1-n_x^4|}
L_{n_{x-}^{14}}^{|n_x^1-n_x^4|} \left( a^2q_x^2/2\right), \label{eq:Cx}
\end{eqnarray}
where $n_i^j$ is the quantum number referring to the $i$-axis of the
particle $j$. We have also used the terms $n_{i+}^{jk}$ and
$n_{i-}^{jk}$, which are defined as $\max(n_i^j,n_i^k)$ and
$\min(n_i^j,n_i^k)$ respectively. The final form for
$C^{\lambda_1}_{\lambda_4}(\bm{q})$ will be
\begin{equation}
C^{n_x^1n_y^1n_z^1}_{n_x^4n_y^4n_z^4} (\bm{q})=  \prod_{i\in\{x,y,z\}}
C^{n_i^1}_{n_i^4}(q_i).
\end{equation}
Using the relation between $D$ and $C$ shown in Eq.~(\ref{eq:D3}), it is
trivial to find out the value of the former convolution integral.

It still remains to calculate the Fourier transform $\tilde{V}(\bm{q})$
of the spherically symmetric interaction potential $V(\bm{r})$.
\begin{equation}
\tilde{V}(q)  =  \sqrt{\frac{2}{\pi}}\frac{1}{q^2} \label{eq:fourier}
\end{equation}
But it will be more convenient to substitute it by
\begin{equation}
\tilde{V}(q)  =  \sqrt{\frac{2}{\pi}}\int_0^{\infty}
e^{-(q_x^2+q_y^2+q_z^2)u}\,\mathrm{d}u\label{eq:fourier2}
\end{equation}

The integration over variables $q_x$, $q_y$ and $q_z$ can be performed
all in the same fashion. Using the symmetry of the problem we only
need to integrate over one variable, i.e. $q_x$ and then use the same
result for $q_y$ and $q_z$. Therefore, integrating over $q_x$ yields:
\begin{eqnarray}
\int_{-\infty}^{\infty}e^{-(u+a^2/2)q_x^2}\left( \frac{aq_x}{2}
\right)^{|n_x^1-n_x^4|+|n_x^2-n_x^3|} \nonumber \\
\times L_{n_{x-}^{14}}^{|n_x^1-n_x^4|}(a^2q_x^2/2)
L_{n_{x-}^{23}}^{|n_x^2-n_x^3|}(a^2q_x^2/2)
\,\mathrm{d}q_x. \label{eq:laguerre1}
\end{eqnarray}
This integral does not vanish if and only if
\begin{equation}
|n_x^1-n_x^4|+|n_x^2-n_x^3| = 2s_x,
\end{equation}
where $s_x=0,1,2,\ldots$. Therefore, using the previous selection rule
and the power series for the associated Laguerre polynomial
\begin{equation}
L_{n}^{l}(x)=\sum_{k=0}^{n}\frac{1}{k!}\binom{n+l}{n-k}(-x)^{k},
\end{equation}
we can write Eq.~(\ref{eq:laguerre1}) as
\begin{eqnarray}
\sum_{k_x=0}^{n_{x-}^{14}}
\frac{(-1)^{k_x}}{k_x!} \binom{n_{x+}^{14}}{n_{x-}^{14}-k_x}
\sum_{k_x=0}^{n_{x-}^{14}}
\frac{(-1)^{k_x^{\prime}}}{k_x^{\prime}!}
\binom{n_{x+}^{23}}{n_{x-}^{23}-k_x^{\prime}} \nonumber \\ 
\times 2^{k_x+k_x^\prime} \left( \frac{a}{2}
\right)^{2s_x+2k_x+2k_x^\prime}  \nonumber \\
\times \frac{(2s_x+2k_x+2k_x^\prime-1)!!}{(2u+a^2)^{s+k_x+k_x^\prime+1/2}}
\sqrt{2\pi}. \label{eq:laguerre2}
\end{eqnarray}

Taking into account only the $u$-dependent part in
Eq.~(\ref{eq:laguerre2}) and its symmetric extension for $y$ and $z$
variables, we end up with the last integral which will lead to the
final result. This last integral is expressed as:
\begin{equation}
\int_{0}^{\infty}\left( 2u+a^2 \right)^{-\Omega-3/2}
\,\mathrm{d}u = \frac{1}{1+2\Omega}\frac{1}{a^{1+2\Omega}},
\end{equation}
where $\Omega=s_x+s_y+s_z+k_x+k_y+k_z+k_x^\prime+k_y^\prime+k_z^\prime$.

Finally, collecting all the terms, we end up with the analytic
expression for the Coulomb interaction matrix elements:
\begin{widetext}
\begin{eqnarray}
\mathcal{V}^{n^1_xn^1_yn^1_zn^2_xn^2_yn^2_z}_{n^3_xn^3_yn^3_zn^4_xn^4_yn^4_z}
& = &  \frac{1}{a} \sqrt{\frac{2}{\pi}}(-1)^{n_x^1 + n_y^1 + n_z^1
+ n_x^4 + n_y^4 + n_z^4 - s_x -s_y - s_z} \nonumber \\
&  & \left(
\frac{2^{n_{x+}^{14}}}{n_{x+}^{14}!} \frac{n_{x-}^{14}!}{2^{n_{x-}^{14}}}
\frac{2^{n_{y+}^{14}}}{n_{y+}^{14}!} \frac{n_{y-}^{14}!}{2^{n_{y-}^{14}}}
\frac{2^{n_{z+}^{14}}}{n_{z+}^{14}!} \frac{n_{z-}^{14}!}{2^{n_{z-}^{14}}}
\right)^{1/2}
\left(
\frac{2^{n_{x+}^{23}}}{n_{x+}^{23}!} \frac{n_{x-}^{23}!}{2^{n_{x-}^{23}}}
\frac{2^{n_{y+}^{23}}}{n_{y+}^{23}!} \frac{n_{y-}^{23}!}{2^{n_{y-}^{23}}}
\frac{2^{n_{z+}^{23}}}{n_{z+}^{23}!} \frac{n_{z-}^{23}!}{2^{n_{z-}^{23}}}
\right)^{1/2} \nonumber \\
& & \sum_{k_x=0}^{n_{x-}^{14}} \frac{(-1)^{k_x}}{k_x!}
\binom{n_{x+}^{14}}{n_{x-}^{14}-k_x}
\sum_{k'_x=0}^{n_{x-}^{23}} \frac{(-1)^{k'_x}}{k'_x!}
\binom{n_{x+}^{23}}{n_{x-}^{23}-k'_x}
\frac{(2s_x+2k_x+2k'_x-1)!!}{2^{2s_x+k_x+k'_x}} \nonumber \\
& & \sum_{k_y=0}^{n_{y-}^{14}} \frac{(-1)^{k_y}}{k_y!}
\binom{n_{y+}^{14}}{n_{y-}^{14}-k_y} \sum_{k'_y=0}^{n_{y-}^{23}}
\frac{(-1)^{k'_y}}{k'_y!} \binom{n_{y+}^{23}}{n_{y-}^{23}-k'_y}
\frac{(2s_y+2k_y+2k'_y-1)!!}{2^{2s_y+k_y+k'_y}} \nonumber \\
& & 
\sum_{k_z=0}^{n_{z-}^{14}} \frac{(-1)^{k_z}}{k_z!}
\binom{n_{z+}^{14}}{n_{z-}^{14}-k_z}
\sum_{k'_z=0}^{n_{z-}^{23}} \frac{(-1)^{k'_z}}{k'_z!}
\binom{n_{z+}^{23}}{n_{z-}^{23}-k'_z}
\frac{(2s_z+2k_z+2k'_z-1)!!}{2^{2s_z+k_z+k'_z}} \nonumber \\
& & \frac{1}{1+2(s_x+s_y+s_z+k_x+k'_x+k_y+k'_y+k_z+k'_z)}.
\label{eq:result}
\end{eqnarray}
\end{widetext}

\section{\label{sec:rec}Recurrence}

Due to the six summatories appearing in Eq.~(\ref{eq:result}), if the
indices start to grow to values say, just of the order of tenths, the
process for calculating a single matrix element can be quite
time-consuming, and thus, a real bottleneck for any numerical
simulation. Using the recurrence relations that the Hermite
polynomials obey, it is possible to find a simple iterative formula
for the matrix elements which will accelerate the process of
calculating the matrix elements.

Let $\{n_-,n_+\}$ be any pair of quantum numbers
$\{n_{i-}^{jk},n_{i+}^{jk}\}$ with $i \in \{x,y,z\}$ and $jk \in
\{14,23\}$, satisfying $n_+ \ge n_-$. Then, the Coulomb matrix
elements will satisfy (remaining indices ommitted for clarity)
\begin{eqnarray}
\mathcal{V}_{n_-}^{n_+} & = &
 \sqrt{\frac{n_++1}{n_-}} \mathcal{V}_{n_--1}^{n_++1}
 + \sqrt{\frac{n_+}{n_-}} \mathcal{V}_{n_--1}^{n_+-1} \nonumber \\
 &  & - \sqrt{\frac{n_--1}{n_-}} \mathcal{V}_{n_--2}^{n_+},
\label{eq:it1}
\end{eqnarray}
for $n_-> 0$. If we consider the unnormalized matrix elements
\begin{eqnarray}
\overline{\mathcal{V}}^{n^1_xn^1_yn^1_zn^2_xn^2_yn^2_z}_{n^3_xn^3_yn^3_zn^4_xn^4_yn^4_z}
& = & \prod_{i\in\{1,2,3,4\}}
(2^{n^i_x}n^i_x!2^{n^i_y}n^i_y!2^{n^i_z}n^i_z!)^{\frac{1}{2}}
\nonumber \\
&  & \times \mathcal{V}^{n^1_xn^1_yn^1_zn^2_xn^2_yn^2_z}_{n^3_xn^3_yn^3_zn^4_xn^4_yn^4_z},
\end{eqnarray}
Eq.~(\ref{eq:it1}) can be transformed to
\begin{equation}
\overline{\mathcal{V}}_{n_-+1}^{n_+} = 
\overline{\mathcal{V}}_{n_-}^{n_++1}
 + 2n_+ \overline{\mathcal{V}}_{n_-}^{n_+-1}
 -2n_- \mathcal{V}_{n_--1}^{n_+}.
\label{eq:it2}
\end{equation}
Another interesting recurrence relation which, this time involves four
indices $\{0,n_+\}$ and $\{m_-,m_+\}$, is the following
\begin{equation}
\overline{\mathcal{V}}_{0,m_-}^{n_+,m_++1} =
\overline{\mathcal{V}}_{0,m_-}^{n_++1,m_+}
+ \overline{\mathcal{V}}_{0,m_--1}^{n_+,m_+}.
\end{equation}

\bibliography{articulo}%

\end{document}